\documentclass[10pt]{iopart}
\usepackage[dvips]{graphicx,epsfig}

\begin{document}
\article[Strange hadron yields and ratios in heavy ion collisions 
at RHIC energy]{}
{Strange hadron yields and ratios in heavy ion collisions at RHIC energy}
\date{30 September 2007}
\author{ G. Hamar$^a$,  L.L. Zhu$^b$, P. Csizmadia$^a$, and P. L\'evai$^a$}
\address{$^a$ RMKI Research Institute for Particle and Nuclear Physics, \\
P.O. Box 114, Budapest, 1525, Hungary}
\address{$^b$ Institute of Particle Physics, Hua-Zhong Normal University, \\
  Wuhan 430079, P.R. China}

\begin{abstract}
Recent experimental data support the presence of quark coalescence
in heavy ion collisions at RHIC energies. Hadronization of 
quark matter and hadron formation in heavy ion collisions can be
described by the coalescence process, and measured data are
reproduced successfully. On the other hand, the theoretical coalescence 
calculations are based on a non-relativistic description. Here we
investigate the robustness of the coalescence description,
using different wave-function overlap during hadron formation. 
\end{abstract}

Quark coalescence has been proposed many years ago
to describe quark matter hadronization~\cite{alcor,alcor2,bialas02}. 
The basic idea was to consider massive constituent quarks in the 
deconfined phase, which are ready to hadronize through 
"coalescence", which is a clustering process driven by an attractive 
force between the properly coloured quark degrees of freedom. 
The presence of these massive excitations in the
quark-matter phase is supported by the analysis of lattice QCD data
and the recognition of massive quasi-particles in the strongly
interacting deconfined phase close to the quark-hadron
phase transition~\cite{PRC98LH}. The attractive force generated by
the gluons (which are considered in this way) is modelled
by non-relativistic colour potential between the quarks and antiquarks.
Mesons are produced by quark-antiquark coalescence. Baryons
are produced in two steps: at first $\overline{3}$ diquarks appear through
quark-quark coalescence, which is followed by a diquark-quark
coalescence into a colourless baryon. 

Although particle yields, ratios and spectra
have been reproduced successfully in ALCOR~\cite{ALCORdat1,ALCORdat2}
and MICOR~\cite{MICOR00,MICORSQM98} coalescence models, 
but thermal models were similarly successful in the low-$p_T$ region 
and more widely used because of their simplicity.
At RHIC energy intense data collection has been performed in the
intermediate-$p_T$ region ($3 < p_T < 8$ GeV/c) and the
measured anomalous proton/pion ratio could have been explained by
quark coalescence and recombination models~\cite{hwa,greco,fries}. 
This fact increased the interest in this microscopical 
hadronization mechanism and more applications appeared. 
The recognition
of valence quark number scaling in the data on asymmetric flow ($v_2$)
supported very strongly the quark matter formation 
and quark coalescence at RHIC and SPS energies~\cite{molnard}. 

The success of the quark coalescence model raised an interesting question,
namely why this model is working successfully, when
it is based on quantum mechanics and 
non-relativistic quark-quark interaction picture.
The main reason is the following: although the quarks are moving
with a relativistic velocity out of the reaction volume, but they
can participate in the coalescence process only if their 
relative velocity is small. In this case quantum mechanics becomes
valid in the comoving system and wave functions start to
play a significant role in the description of hadron formation.

In this paper we investigate the sensitivity of the hadron yields
on the use of the different quark and hadron wave functions.
We will consider plain wave and gaussian wave functions for quarks
and antiquarks, where the gaussian choice indicates the presence of
a finite volume deconfined quark matter. For mesons we use
the same two choices and extend the list with the hydrogen-like
wave function, which is connected to an analogy between
electromagnetic and strong forces with proper coupling constant. 
For diquarks we use the same three choices. However, the diquark
and the plasma state may interact and the diquark wave function
can be modified, flipping between the plain wave, gaussian and
hydrogen like wave functions. Thus baryons are produced
similarly to mesons, but in two steps, as we will discuss it in details. 

At first we determine the quark coalescence cross section in quantum 
mechanics. We use the quantum mechanical pick-up reaction~\cite{Schiff}:
$q_1+Q\vert_{q_2} \longrightarrow h+Q'$, where 
quark $q_1$ picks up quark $q_2$ from the plasma $Q$,  
producing prehadron $h$ and plasma $Q'$. 
The quantum mechanical amplitude of the pick up reaction  
is given:
\begin{eqnarray}
g_{gh} &=& V_g \frac{-M_{h,Q'}}{2 \pi} \int d^3\vec{x}_1 d^3\vec{x}_2 \   
\widetilde{\Psi}^*(\vec{x}_1,\vec{x}_2) V(\vec{x}_1-\vec{x}_2) 
\phi_1(\vec{x}_1) \phi_2(\vec{x}_2) \, , \label{ggh}
\end{eqnarray}
where $\phi_i(\vec{x}_i)$ is the wave function of the $q_i$ quark, 
and $\widetilde{\Psi}(\vec{x}_1,\vec{x}_2)$ belongs to the
prehadron with mass $M_h$.
$M_{h,Q'}$ is the reduced mass of $h$ and $Q'$. 
Since $M_h \ll M_{Q'}$ therefore $M_{h,Q'} \simeq M_h$.
The standard two-body coordinates can be introduced as
relative distance vector (${\vec r}$), 
relative momentum vector (${\vec k}$), 
space and momentum vector of the center of mass (${\vec X}$, ${\vec P}$).
Since the outgoing prehadron has the momentum 
$\vec{P}=\vec{p}_1+\vec{p}_2$, thus the wave function $\widetilde{\Psi}$ 
is simplified to  $\Psi({\vec r}) \cdot e^{i\vec{P}\vec{X}}$. 

In eq.(\ref{ggh})  $V(\vec{x}_1-\vec{x}_2)$ denotes the
quark-quark interaction. Here we introduce the 
Yukawa-potential, which depends on the relative distance, $r$:
\begin{equation}
V(r)=-{\alpha}{\langle \lambda_i\lambda_j \rangle} \frac{e^{-m_g r}}{r}
\end{equation}
The screening mass has a relatively large value at $T\approx T_c$:
$m_g=0.8$ GeV~\cite{PRC98LH}. The colour factor 
${\langle \lambda_i\lambda_j \rangle}$ is determined by the colour
combination of the interacting particles.
In the limit of $m_g=0$ Coulomb potential is restored,
which has been used in the ALCOR model~\cite{alcor}.
In the MICOR model a Yukawa potential has been considered~\cite{MICOR00}.

In $2\rightarrow 2$ reactions (e.g. $a + b \to c +d $) the cross section
can be determined from the amplitude as
\begin{eqnarray} 
\sigma (k) &=& \frac{v_{cd}}{v_{ab}} \ {\left| g_{gh}(k) \right|}^2  \ ,
\label{sigk}
\end{eqnarray}
where $v_{ab}$ and $v_{cd}$ are the relative velocities.
In our case the factor $v_{cd}$ could be large,
because it is the relative velocity of the outgoing (pre)hadron 
and the plasma. Thus we have two choices during evaluation of
eq.~(\ref{sigk}):
\begin{eqnarray}
 \sigma(k) v_{q_1 q_2} \vert_{classical}  &=& 
\frac{P}{M_h} \ {\left| g_{gh}(k) \right|}^2    \\
 \sigma(k) v_{q_1 q_2} \vert_{relativistic} &=& 
\frac{P}{\sqrt{P^2 + M_h^2}} \ {\left| g_{gh}(k) \right|}^2
\label{sigrel}
\end{eqnarray}

Now we are ready to
introduce quark and prehadron wave functions to calculate
numerically applicable coalescence cross sections from
eqs.~(\ref{ggh})-(\ref{sigrel}).
In the lack of precise quark wave functions, we will use
simple functions, which will be displayed and discussed
after introducing the calculation of hadron production
rates and yields.

Previously, the coalescence cross sections were obtained from 
quantum mechanics.
The wanted hadron production rates and yields can be 
determined by a statistical method based on 
rate equations using the former cross sections. 

Prehadron $h$ is composed from quarks $q_1$ and $q_2$ via coalescence,
and its production is proportional to the densities of the constituents,
$n_1$ and $n_2$~\cite{alcor,MICOR00}:
\begin{eqnarray}
 \partial _\mu (n_h u^\mu)  = 
 \langle \sigma^h_{12} v_{12} \rangle \ n_1 \, n_2 \ .
\end{eqnarray}

The rate, $\langle \sigma^h_{12} v_{12} \rangle$, is calculated
as a phase space average:
\begin{eqnarray}
\langle \sigma^h_{12} v_{12} \rangle &=& 
\frac{\int d^3 \vec{p}_1 d^3 \vec{p}_2 d^3 \vec{x}_1 
d^3 \vec{x}_2 \rho_{12}(\vec{x}_1, \vec{x}_2)  
f_1(\vec{x}_1, \vec{p}_1) f_2(\vec{x}_2, \vec{p}_2)  
\sigma v_{12}}{\int d^3 \vec{p}_1 d^3 \vec{p}_2 d^3 
\vec{x}_1 d^3 \vec{x}_2 \rho_{12}(\vec{x}_1, 
\vec{x}_2)f_1(\vec{x}_1, \vec{p}_1) f_2(\vec{x}_2, \vec{p}_2)} 
\label{aver0}
\end{eqnarray}
where $f_i (\vec{x}_i, \vec{p}_i)$ are the quark momentum distributions 
and $\rho$ describes the locality of the quark coalescence. 
Requiring that quarks with the same location are able to coalesce, 
$\rho$ becomes a simple Dirac delta. Assuming
isotrop plasma state,  eq.(\ref{aver0}) is simplified into the following 
expression:
\begin{eqnarray}
\langle \sigma^h_{12} v_{12} \rangle &=& \frac{\int d^3 \vec{p}_1 d^3\vec{p}_2  
\ f_q(m_1, \vec{p}_1) f_q(m_2, \vec{p}_2)  \sigma v_{12}}
{\int d^3 \vec{p}_1 d^3 \vec{p}_2 \ f_q(m_1,\vec{p}_1)f_q(m_2,\vec{p}_2)} 
\end{eqnarray}

During the evaluation of the rate one can use any proper distribution function.
Because of the massive quarks we can
use non-relativistic Boltzmann
distribution:
\begin{eqnarray} f_q^{Boltzmann}(m, \vec{p})&=&e^{-\frac{p^2}{2mT}} 
\end{eqnarray}

In parallel,  relativistic quark distribution can be used also,
namely  J\"uttner distribution, which is simplified in the
local rest frame of the expanding fireball:
\begin{eqnarray} 
f_q^{Juttner}(m, \vec{p})&=&e^{-\frac{u_\mu p^\mu}{T}}=
e^{-\frac{\sqrt{p^2+m^2}}{T}} 
\end{eqnarray}
We will use both distributions and investigate the sensitivity of the
rate on this choice and the temperature $T$.
Furthermore, in the above quark momentum distribution functions
we will use quark masses $m_q=0.3$ GeV and $m_s=0.5$ GeV,
which values are verified in the analysis of the lattice data
close to the quark-hadron phase transition~\cite{PRC98LH}.

Now we have all expression to investigate prehadron
production from quark matter. If we define the necessary
wave functions for quarks and prehadrons, then we can perform 
numerical calculations and determine the wanted particle productions.
As we mentioned earlier, we introduce two types of
wave functions for quarks:
\begin{eqnarray}
\phi_i &= \frac{1}{\sqrt{V_q}} e^{i \vec{p}_i \vec{x}_i} 
\ \ \  &\hspace{1truecm} {\rm PW:} \ {\rm plain \ wave }  \label{qpw} \\
\phi_i &= \frac{1}{(2 \pi \rho^2)^{{3}/{4}}} e^{- \frac{{x_i}^2}{4 \rho^2}}
          e^{i \vec{p}_i \vec{x}_i} 
\ \ \  &\hspace{1truecm} {\ \ \ \rm G:} \ {\rm gaussian  } \label{qg} 
\end{eqnarray}
For prehadrons we introduce the following wave functions:
\begin{eqnarray}
{\widetilde \Psi} &= \frac{1}{\sqrt{V_f }}  \frac{1}{\sqrt{V_h}}
\ e^{i \vec{P} \vec{X}} 
\ \ \  &{\rm PW:} \ {\rm plain \ wave }  \label{hpw} \\
{\widetilde \Psi} &= \frac{1}{\sqrt{V_f }}
 \frac{1}{\sqrt{\pi a^3}} \   e^{- \frac{r}{a}}
        \   e^{i \vec{P} \vec{X}}  
\ \ \  &{\ \ \ \rm H:} \ {\rm hydrogen \ like  }  \label{hh}  \\ 
{\widetilde \Psi} &= \frac{1}{\sqrt{V_f }}
 \frac{1}{(2\pi \eta^2)^{{3}/{4}}} \  e^{- \frac{r^2}{4 \eta^2}}
       \    e^{i \vec{P} \vec{X}} 
\ \ \  &{\ \ \ \rm G:} \ {\rm gaussian  }  \label{hg}
\end{eqnarray}
These expressions contain the volume normalization factors
related to the characteristic volume of quarks ($V_q$),
prehadrons ($V_h$), and the fireball volume ($V_f$).

\begin{figure}
\begin{center}
\includegraphics[width=6.45cm]{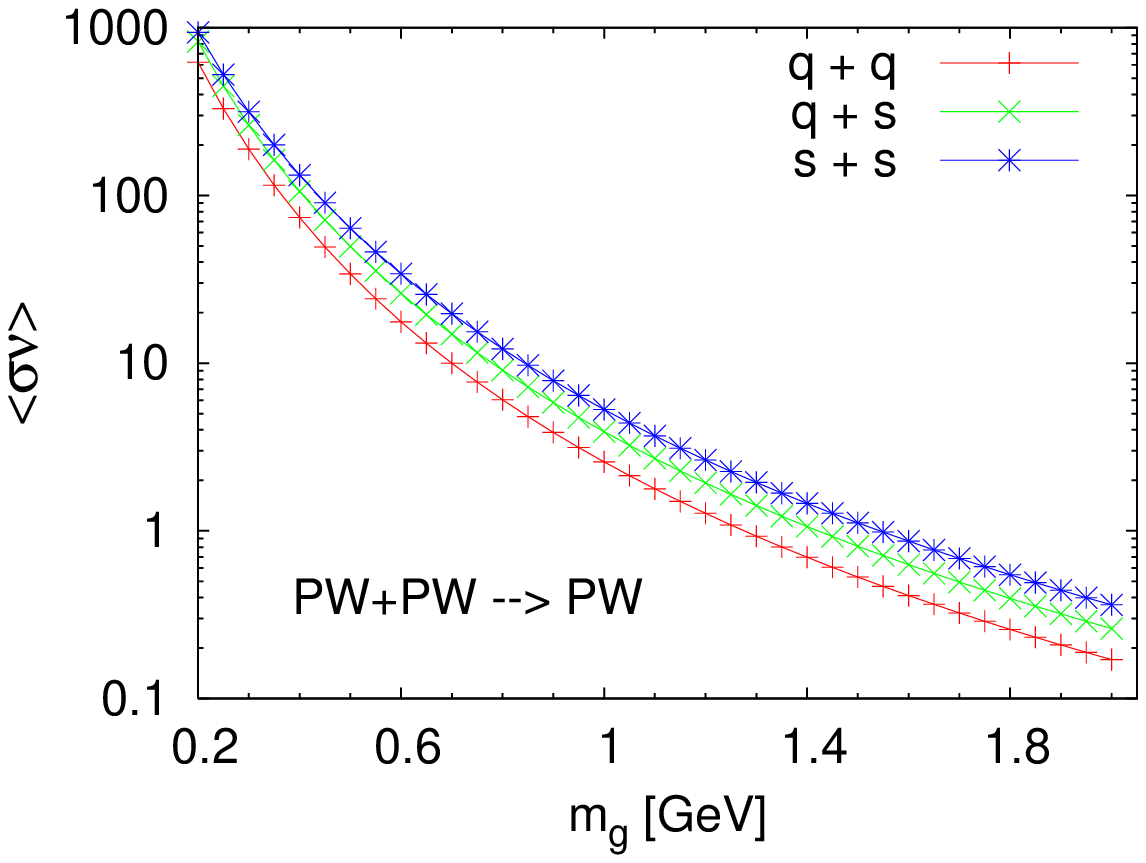}
\includegraphics[width=6.45cm]{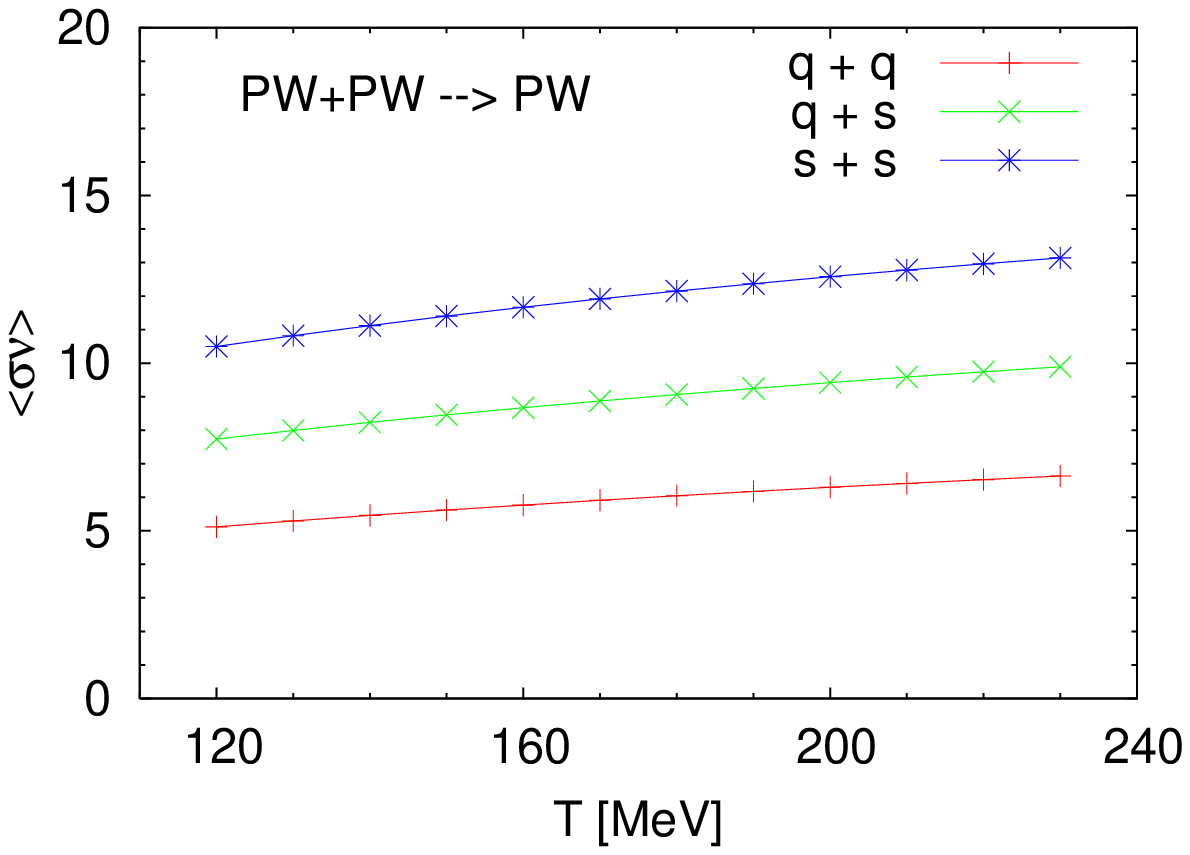}
\end{center}
\caption{\protect\label{mgt} The dependence of the coalescence rate
($\langle \sigma v \rangle$)
on gluon mass ($m_g$) at fix temperature ($T=180$ MeV) (left) 
and on temperature (T) at fix gluon mass ($m_g=800$ MeV) (right)
for the case of quark plain waves coalesce into mesonic plain wave.
} \vspace*{0.5cm}
\end{figure}

The wave functions in eqs.(\ref{qpw})-(\ref{qg}) and
(\ref{hpw})-(\ref{hg}) offer us 9 combinations, which can be doubled
by the application of Boltzmann and  J\"uttner momentum distributions.
For consistency we will calculate meson and baryon productions in the
same (fixed) wave function combination.
We note that ALCOR model~\cite{alcor} used the $PW+PW \rightarrow H$
combination, MICOR model~\cite{MICOR00} applied the
$PW+G \rightarrow H$ choice, including a gaussian localization
for the picked-up quark. 

In the case of process $PW+PW \rightarrow PW$   
averaging with the Boltzmann distribution (named as case 'B1'),
one can obtain a compact
expression for the rate:
\begin{equation}
\langle \sigma^h v \rangle_{B1} =
\, \frac{ V_g^2 V_t}{V_q^2 V_h} 
\frac{M_h (m_1 + m_2)^2}{(m_1 m_2)^{{3}/{2}}}
\frac{\alpha_s^2}{ \pi} \frac{1}{T}
\int dk \ 
\frac{k^2 e^{-\frac{m_1+m_2}{T \, m_1 m_2} \, k^2} }{(k^2+{m_g}^2)^2}
\ . \label{sigmv0}
\end{equation}
This expression shows that the volume terms can be collected into
a prefactor together with the coupling constant $\alpha_s$,
which is valid for every combinations.
This prefactor should be fitted from 
one data point, and all other particle yields become calculable.
On the other hand, this factor will drop out from particle ratios 
(calculated with the same wave function combination). 
In the remaining 17 combinations much longer
expressions appear for the rates, but all of them can be calculated 
numerically. 

In our investigation 
we recognized that combinations containing at least one quark plain wave
resulted numerically different values, but they differed in a 
constant factor, only.
This is the reason why ALCOR's and MICOR's results on particle yields
are generally the same, although different wave function sets have been 
used.

Now we investigate 
the sensitivity of the coalescence rate on the mass parameter 
of the Yukawa potential ($m_g$) and the temperature (T).
In Figure 1 we display the calculated rates (with an arbitrary scale)
in the case of $PW+PW \rightarrow PW$, using 
the expression of eq.~(\ref{sigmv0}).
One can see (left) that the meson production rates 
strongly depend on the gluon mass at fix temperature (here we
choose $T=180$ MeV). 
Considering the temperature dependence (right),
the rates are very much insensitive on this parameter at
fix gluon mass ($m_g=800$ MeV). 
This fact verifies the applicability of coalescence models
for cross-over phase transitions, where no sharp transition
temperature exists. 
In Figure 1
one can recognize the presence of a flavour dependence,
which is connected to the heavier mass of the strangeness.

\begin{figure}
\begin{center}
\includegraphics[width=6.45cm]{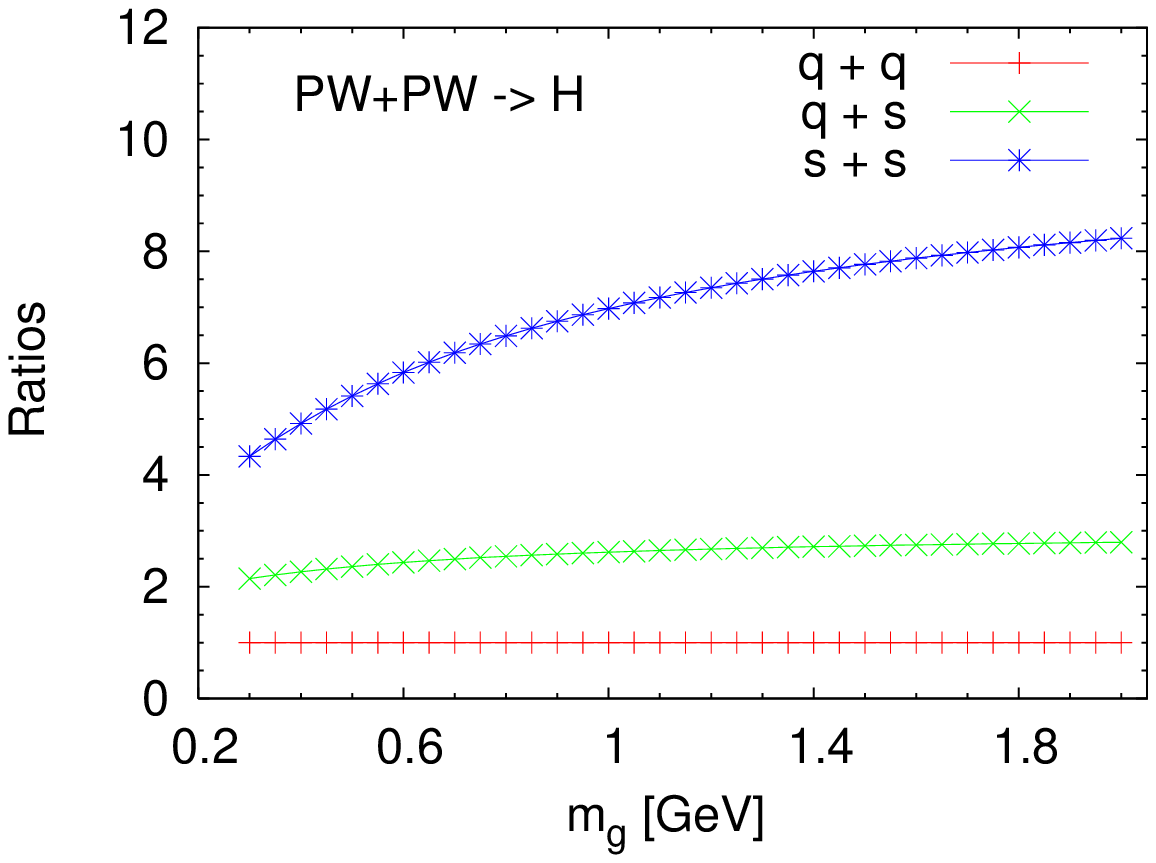}
\includegraphics[width=6.45cm]{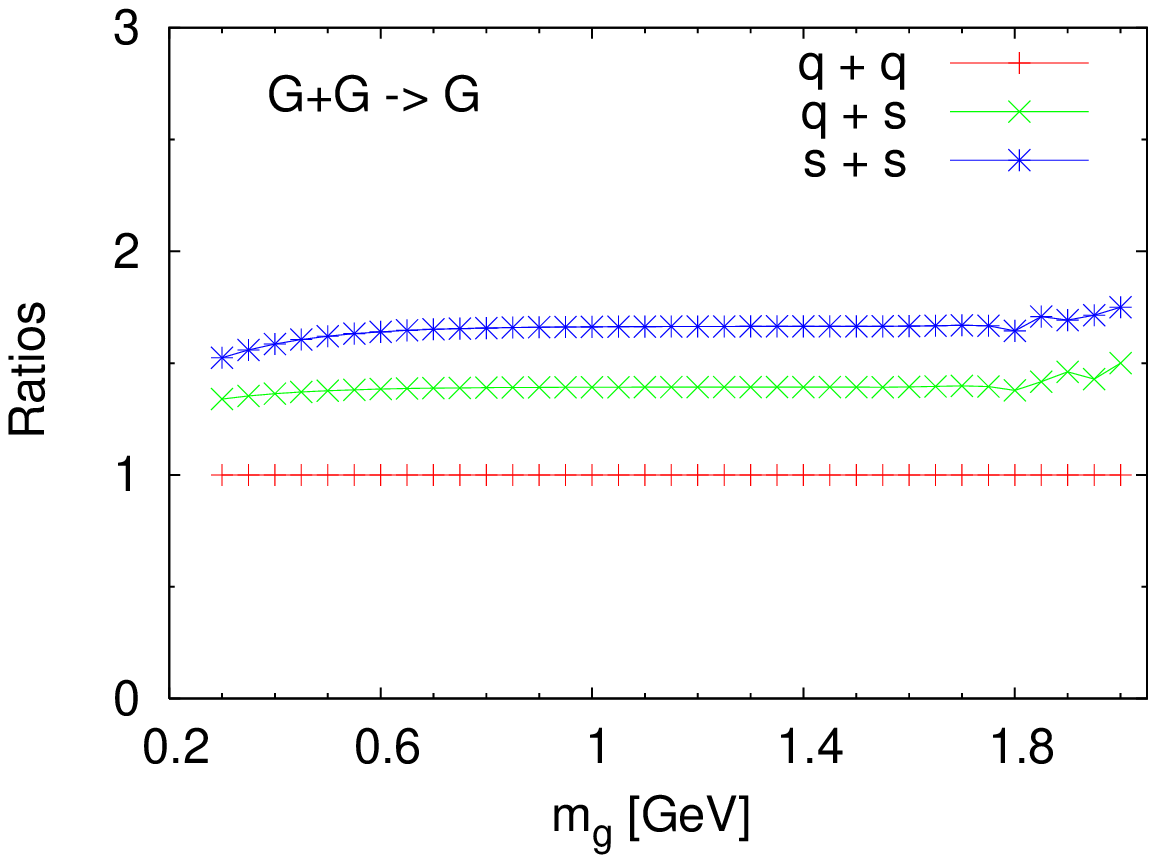}
\end{center}
\caption{\protect\label{mgtr} The gluon mass ($m_g$)
dependence of the production ratios of mesons with different flavour
combinations, relative to the light meson
at fix temperature ($T=180$ MeV).
We display the numerical results for the
$PW+PW \rightarrow H$ (left) and
$G+G \rightarrow G$ (right). 
} \vspace*{0.5cm}
\end{figure}

This flavour dependence is investigated further through the
particle ratios, in which case uncertainties connected
to unknown volume factors disappear. 
The temperature dependence of the ratios is negligible,
as one could expect from Figure 1.  We have found that
the strong gluon mass dependence drops out for ratios
in all 18 wave function combinations. 
In Figure 2 we show our numerical results
for the case of $PW+PW \rightarrow H$ (left)
and the case of $G+G \rightarrow G$ (right),
which results illustrate the weak dependence of the
ratios on the gluon mass in very different cases.

These results indicate and prove the robustness of the coalescence model
within a fixed wave function combination, since the theoretical
results depend very weakly on temperature and/or gluon mass.
This feature nicely support the applicability of coalescence models
to describe cross-over phase transition, which is expected to happen
in a wide temperature region.

Now we would like to investigate the sensitivity of the coalescence
model on different wave function setups.
For this task we want to use real data measured at RHIC energy.
Our method is the following: we use one measured meson
ratio to fix the open parameters, recalculate other 
measurable hadron ratios, and investigate the difference
between the obtained results in difference wave function setups. 
One of our candidate for the starting point
is the ratio $\Phi/K^*=0.60 \pm 0.15$ measured at RHIC in central
Au+Au collisions at $\sqrt{s} = 200$ AGeV~\cite{RHICPhiK}.
(We use the middle value and neglect the influence of the 
error.)  
In the coalescence model this ratio appears in the following way:
\begin{equation}
\frac{N_\phi}{N_{K^*}}=
\frac{\langle \sigma^h v \rangle_{ss} \cdot N_s N_s}
     {\langle \sigma^h v \rangle_{sq} \cdot N_s N_q}  
\label{fi_per_k}
\end{equation}
The rates in eq.~(\ref{fi_per_k}) can be calculated in a fixed
wave function setup. Thus we can determine the missing $N_s/N_q$
factor from the measured ${N_\phi}/{N_{K^*}}$ value.
Then different strange-non-strange meson and baryon ratios can
be calculated from the model. Here we will calculate the ratios
$K^*/\rho^0$, $\Sigma^*/\Delta$, $\overline{\Xi}^*/\Sigma$
and $\Omega/\overline{\Xi}^*$. These resonances have been
measured~\cite{RESON} and the above ratios could be determined from
existing experimental data.

Table 1. summarizes our numerical results. We can see that different
wave function setups result very different $N_s/N_q$ values,
namely 100-150 \% difference can be seen 
between the smallest and the largest values of quark ratios. 
On the other hand, this uncertainty drops
to a 10-15 \% difference, 
both for mesonic and baryonic ratios.
These results prove most strikingly why the coalescence models
yield very good agreement during data reconstruction, if we
start from one measured values.
Further analyses are in progress to reveal a deeper 
connection between the manifestation of conservation laws
and the structure of the quantummechanical description of quark coalescence.

\begin{table}
\caption{
Hadron ratios in different wave function setups with Boltzmann (upper part)
and J\"uttner distributions (lower part). 
The ratio $N_s/N_q$ is determined from the
fixed ratio $\Phi/K^{*}=0.6$ measured at RHIC~\cite{RHICPhiK}.
Other particle ratios are calculated from the corresponding wave 
function setup, using the former strange to light quark ratio.}
\begin{indented}
\item[]\begin{tabular}{@{}l|l|llll}
\br
{\bf Model} &  $N_s/N_q$  & $K^{*}/\rho^0$  &$\Sigma^{*}/\Delta$ & 
$\Xi^{*}/\Sigma$ & $\Omega/\Xi^{*}$ \\
\br
PW+PW $\rightarrow$ PW & 0.448 & 0.671 & 0.790 & 0.700 & 0.644 \\
PW+PW $\rightarrow$ H  & 0.235 & 0.598 & 0.848 & 0.785 & 0.746 \\
PW+PW $\rightarrow$ G  & 0.474 & 0.686 & 0.791 & 0.690 & 0.628 \\
G+G   $\rightarrow$ PW & 0.441 & 0.668 & 0.790 & 0.702 & 0.648 \\
G+G   $\rightarrow$ H  & 0.227 & 0.593 & 0.849 & 0.791 & 0.753 \\
G+G   $\rightarrow$ G  & 0.503 & 0.700 & 0.791 & 0.681 & 0.614 \\
\br
PW+PW $\rightarrow$ PW & 0.401 & 0.701 & 0.857 & 0.721 & 0.644 \\
PW+PW $\rightarrow$ H  & 0.206 & 0.624 & 0.928 & 0.815 & 0.752 \\
PW+PW $\rightarrow$ G  & 0.424 & 0.706 & 0.843 & 0.708 & 0.631 \\
G+G   $\rightarrow$ PW & 0.396 & 0.701 & 0.862 & 0.724 & 0.647 \\
G+G   $\rightarrow$ H  & 0.201 & 0.620 & 0.931 & 0.819 & 0.757 \\
G+G   $\rightarrow$ G  & 0.435 & 0.712 & 0.845 & 0.706 & 0.627 \\
\br
\end{tabular}
\end{indented}
\end{table}

This work has been supported in part by 
the Hungarian OTKA under grants No. NK062044, IN71374, K67942 and 
the National Natural Science Foundation of
China under grant No 10475032.


\vspace{0.2truecm}

\begin{thebibliography}{99}

\bibitem{alcor}
        T.S. Bir\'o, P. L\'evai, and J. Zim\'anyi,
          Phys. Lett.  {\bf B347}, 6 (1995); Phys. Rev. {\bf C59}, 1574 (1999).

\bibitem{alcor2}
        T.S. Bir\'o, T. Cs\"org\H o, P. L\'evai, and J. Zim\'anyi,
          Phys. Lett.  {\bf B472}, 243 (2000).

\bibitem{bialas02}
        A. Bialas, Phys. Lett. {\bf B532}, 249 (2002);
        {\it ibid.} {\bf B579}, 31 (2004).

\bibitem{PRC98LH}
        P. L\'evai, U. Heinz,
        Phys. Rev. {\bf C57}, 1879 (1998).

\bibitem{ALCORdat1}
        T.S. Bir\'o, P. L\'evai, J. Zim\'anyi,
        J. Phys.    {\bf G25}, 311 (1999); 
        {\it ibid.} {\bf G27}, 439 (2001);
        {\it ibid.} {\bf G28}, 1561 (2002);
        {\it ibid.} {\bf G31}, 711  (2005).

\bibitem{ALCORdat2}
        J. Zim\'anyi, T.S. Bir\'o, T. Cs\"org\H o, P. L\'evai, 
        Phys. Lett. {\bf B472}, 243 (2000).

\bibitem{MICOR00}
P. Csizmadia and P. L{\'e}vai, 
Phys. Rev. {\bf C61}, 031903 (2000);
Acta Phys. Hung. {\bf A22}, 371 (2005);
{\it ibid.} {\bf A27}, 433 (2006).

\bibitem{MICORSQM98}
P. Csizmadia, P. L\'evai, S.E. Vance, T.S. Bir\'o, M. Gyulassy,
and J. Zim\'anyi, J. of Phys. {\bf G25}, 321 (1999).

\bibitem{hwa}
        R.C. Hwa and C.B. Yang,
          Phys. Rev. {\bf C66}, 064903 (2002).

\bibitem{greco}
        V. Greco, C.M. Ko, P. L\'evai,
           Phys. Rev. Lett. {\bf 90}, 202302 (2003);
           Phys. Rev. C {\bf 68}, 034904 (2003).

\bibitem{fries}
        R.J. Fries, B. M\"uller, C. Nonaka, S.A. Bass,
           Phys. Rev. Lett. {\bf 90}, 202303 (2003);
           Phys. Rev. {\bf C68}, 044902 (2003).

\bibitem{molnard}
        D. Moln\'ar, S.A. Voloshin,
           Phys. Rev. Lett. {\bf 91}, 092301 (2003);
        Z.W. Lin, D. Molnar,
           Phys. Rev. {\bf C68}, 044901 (2003).

\bibitem{Schiff}
        Rearrangement collision, Section 34.,
        in L.I. Schiff:
        Quantum Mechanics, Second edition, McGraw Hill, New York, 1955.

\bibitem{RHICPhiK}
        J. Adams {\it et al.}, 
        Phys. Rev. {\bf C71}, 064902  (2005).

\bibitem{RESON}
       B.I. Abelev {\it et al.},
       Phys. Rev. Lett. {\bf 97}, 132301 (2006).






\end{thebibliography}
\end{document}